\documentclass[10pt,a4paper]{iopart}
\usepackage[compress]{cite} 
\usepackage{graphicx}
\usepackage{amssymb}
\usepackage{multicol}
\usepackage{xcolor}
\usepackage{soul}

\begin{document}

\title[Dielectric antenna effects in piezoelectric sensors]{Dielectric antenna effects in integrating line piezoelectric sensors for optoacoustic imaging}
\author{R. M. Insabella$^1$, M. G. Gonz\'alez$^{1,2}$\footnote{Corresponding author.}, E. O. Acosta$^1$, G. D. Santiago$^1$}
\address{$^1$ Universidad de Buenos Aires, Facultad de Ingenier\'ia, GLOmAe, Paseo Colon 850 (C1063ACV), Buenos Aires, Argentina}
\address{$^2$ CONICET, Godoy Cruz 2290 (C1425FQB), Buenos Aires, Argentina.}
\ead{mggonza@fi.uba.ar}

\begin{abstract}
This work studies the adverse effects, as regards noise, of immersing in water an integrating line piezoelectric detector devoted to optoacoustic imaging. We found that the sensor, in conjunction with the acoustic coupling medium (water), behaves as a resonant dielectric antenna. This phenomenon limits the performance of the system because it efficiently captures unwanted electromagnetic signals. The requirement of good acoustic coupling between the water and the sensor precluded the use of a standard metallic shielding enclosure. Therefore, we resorted to a silver-paint based electrical shield deposited on the detector. This easy-to-implement and low-cost solution significantly increases the signal to noise ratio and does not degrade the acoustic performance. The noise reduction allows the use of a better transimpedance amplifier with higher gain and bandwidth; thus achieving a very sensitive, low-noise detection system.
\end{abstract}

\vspace{2pc}
\\
\\
%
\submitto{\MST}
%
\maketitle
%
\ioptwocol

\section{Introduction}
\label{s:intro}

The  optoacoustic (OA) phenomenon is the generation of acoustic waves due to thermoelastic expansion caused by absorption of short optical pulses. When the OA technique is used to perform tomography, the pressure profile generated by the optical excitation is captured with ultrasonic sensors that surround the area of interest. An OA configuration that obtains images consists of three basic elements: a light source, a detection setup to measure the acoustic waves and a system that processes the recorded signals. Generally, the detection system consists of one or more ultrasonic transducers placed in a vessel filled with water, which serves as a coupling medium between the sample and the sensor \cite{paltauf2017,bauer2017,jiang2019,brecht2009}. In OA tomography (OAT), it is possible to classify these devices into two categories: piezoelectric (or capacitive) transducers, in which the measured electrical signal is directly proportional to the pressure; and optical detectors, which are sensitive to changes in the optical path length, generated by pressure waves \cite{lutzwieler2013}. 

Sensors based on piezoelectric technology are widely used when high sensitivity and low cost are sought \cite{paltauf2017}. Thin-film, polymer piezoelectric transducers are usually employed because these sensors respond well to OAT wideband signals with low amplitude \cite{brown2000}. Even though point sensors offer optimal resolution \cite{vidal2017}, small piezoelectric transducers with an active size in the range of micrometers are difficult to manufacture and suffer from poor sensitivity \cite{paltauf2009}. One way of overcoming the resolution problem with finite size transducers relies on large area, integrating detectors and further processing to take into account the detector's characteristics in the image reconstruction\cite{burgholzer2007}. However, a larger area leads to increased sensor capacitance, an undesirable feature because it complicates the design of the amplifying stages \cite{gonzalez2016}. Integrating line sensors are a good trade-off between sensitivity and capacitance \cite{nuster2009,burgholzer2007,paltauf2017,abadi2018} and also have other advantages: i) reduction of the imaging problem to two dimensions (simpler reconstruction algorithms)\cite{burgholzer2007}, ii) the drop of signal amplitude is proportional to the inverse of the square root of the distance \cite{paltauf2017}; and iii) its implementation is relatively easy \cite{paltauf2009}. 

In this work, we study the adverse effects, as regards noise, of immersing in water an integrating line piezoelectric detector devoted to optoacoustic imaging. This type of sensors, immersed in a water-filled vessel, resembles a dielectric resonator antenna (DRA). Due to the high relative dielectric permittivity of water ($\epsilon_{r}=80$), this antenna has a resonance frequency much lower than a similar one in air and the detector becomes an efficient collector of electromagnetic noise at frequencies in the same range of the OA signals. In such a situation, the electromagnetic shielding becomes more important but a standard, fully enclosing Faraday cage can not be used for it would also attenuate the acoustic wave. We studied this problem in home-made ultrasonic sensors and the results show that an electrical shield based on conductive silver paint significantly reduces the noise produced by the DRA effect and allows the use of a better amplifier with higher gain and bandwidth. This way, a very sensitive, low-noise detection system can be achieved.   

The paper is organized as follows. In Sect.~\ref{s:charac} we study the DRA effect in line integrating piezoelectric sensors. In addition, we detail the materials and methods used to characterize detection systems based on different sensors and transimpedance amplifiers. In Sect.~\ref{s:toa} we compare the performance of sensors with and without electrical shield. To achieve this we compare the responses in a bidimensional OAT and we measure the system resolution. Finally, we present the conclusions in Sect.~\ref{s:conclu}.  

\section{DRA effect in piezoelectric sensors}
\label{s:charac}

\subsection{Home-made line piezoelectric detectors}

\begin{figure}
	\includegraphics[width=\columnwidth]{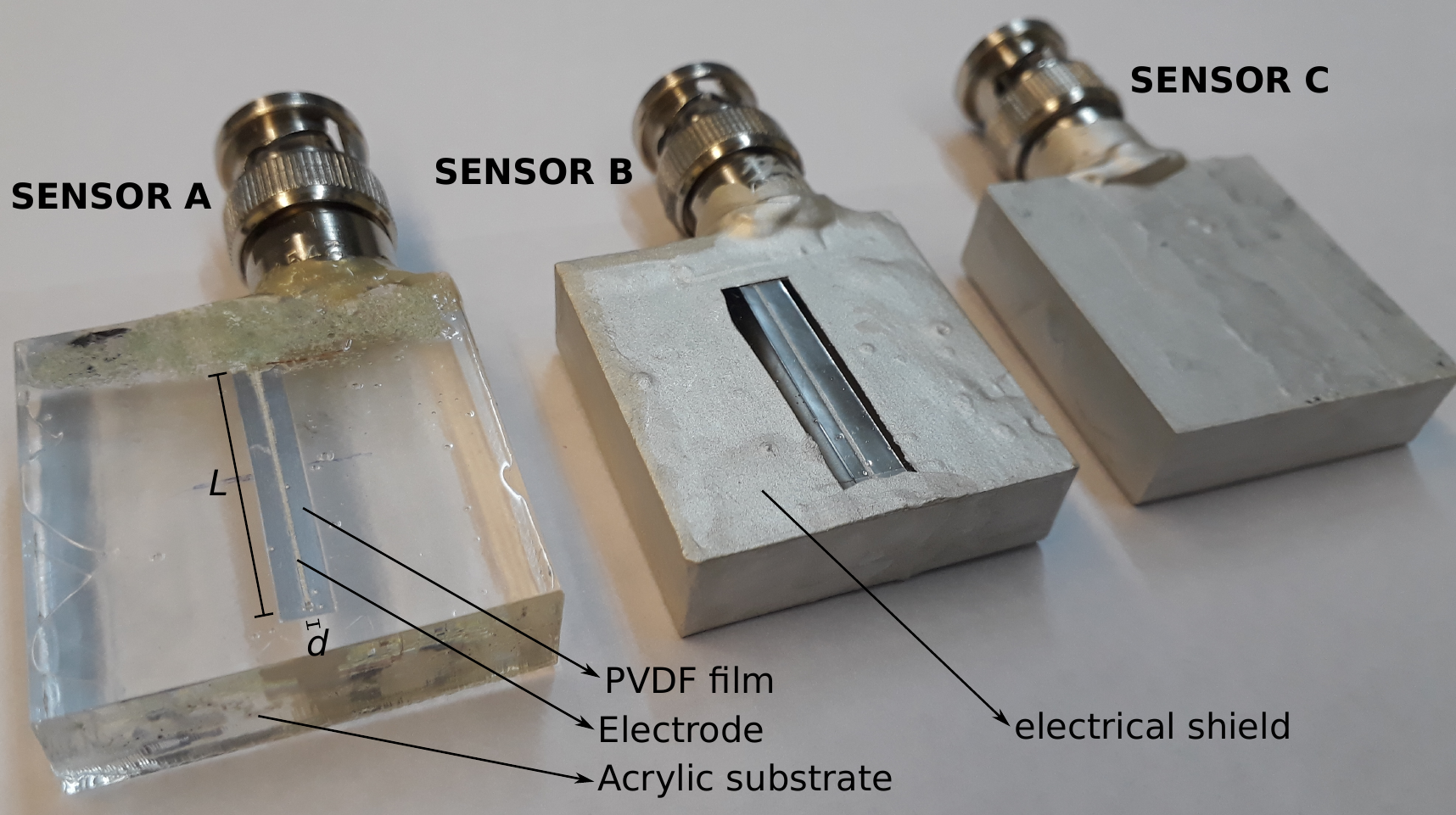}
	\caption{Picture of the polymeric piezoelectric sensors studied in this work}
	\label{fig:sensors}
\end{figure}

Fig.~\ref{fig:sensors} shows a picture of the polymeric piezoelectric sensors studied in this work. Sensor A consists of a PVDF film (25 $\mu$m thickness) with two electrodes attached to an acrylic substrate with dimensions 30 mm x 30 mm x 10 mm. One of the film electrodes, the one in contact with the substrate, was an aluminum foil. The other electrode was made with silver paint, achieving an active detection area of approximately 0.7 mm($d$) x 24 mm ($L$). The shape and size of the transducer were chosen based on the following criteria: i) the active area should resemble a line; ii) maximum sensitivity and minimum capacitance are sought. The latter is very important because the sensor's capacitance strongly influences the frequency response and bandwidth of the detection system \cite{gonzalez2016}. To protect the PVDF film and the silver-painted electrode from the water (to avoid changes in their dielectric properties), we covered it with non-conductive, acrylic paint ( thickness $<$ 1 mm). This paint adheres well to the sensor (keeping it dry) and its acoustic impedance ($c$=1430 m/s and $\delta$=1190 kg/m$^3$ at 25 $^o$C) is very similar to that of water at the frequency range of interest of this work. Moreover, as it was described in a previous work \cite{abadi2018}, we verified that this layer does not introduce a significant reduction of the sensitivity nor any harmonic distortion. 

Sensors B and C were implemented following the same method mentioned above, but with an electrical shield based on conductive silver paint. It can be seen, in Fig.~\ref{fig:sensors}, that sensor B has a window without paint around the active detection area, whilst sensor C is fully covered with paint. Even though the role and convenience of shielding are known, their relevance in this case will be addressed in detail.

\subsection{Detection system characterization}

For the electrical characterization, we used the method described in ref. \cite{gonzalez2016}. Firstly, we measured the sensor's capacitance in the range from 10 Hz to 10 MHz with a bridge circuit driven by a synthesized signal generator. The results showed the three sensors have identical capacitance (60 pF at 10 kHz) and a dependence on the frequency that fits well with the Havriliak-Negami function \cite{gonzalez22014}. Secondly, the frequency response of the detection system (sensor + amplifier) was determined with a network analyzer at frequencies up to 200 MHz. We tested two different transimpedance amplifiers (TIA), EG$\&$G Optoelectronics Judson PA-400 (TIA1) and FEMTO HCA-100MHz-50K-C (TIA2) with flatband gains of 4 k$\Omega$ and 50 k$\Omega$, respectively. The detection system which combines sensor A and TIA1 has a frequency cut-off (-3 dB) of 20 MHz and 63 MHz with TIA2. Similar results ($\pm$ 5 $\%$) were measured with sensors B and C. 

\begin{figure}
	\includegraphics[width=\columnwidth]{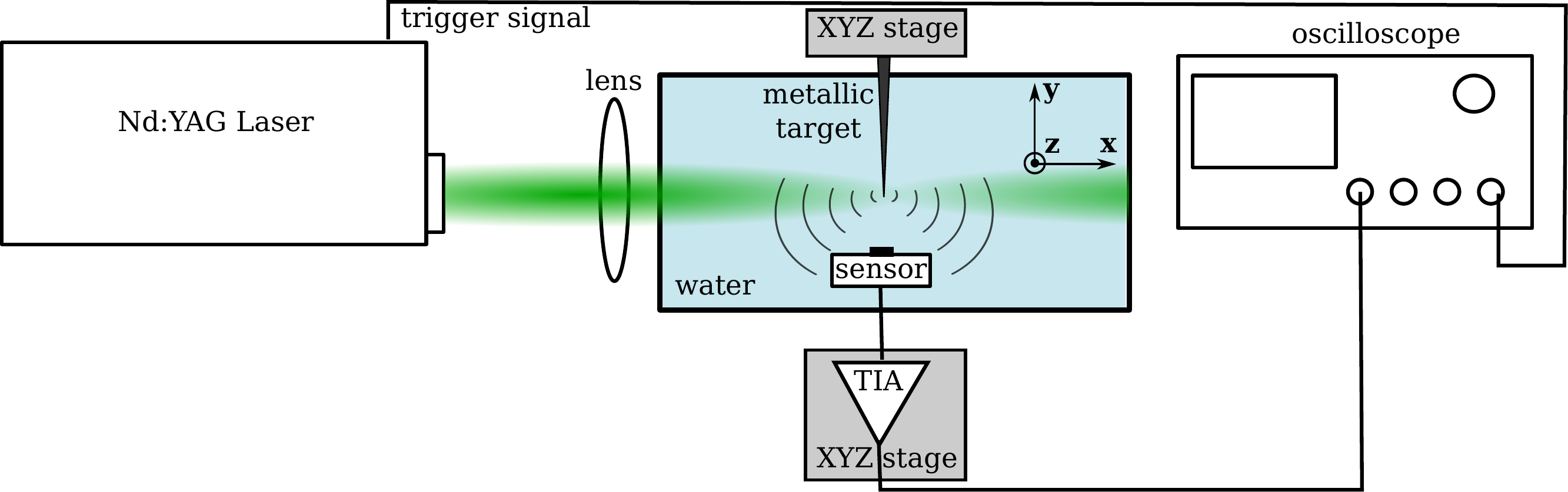}
	\caption{Experimental setup for the acoustic characterization of the sensors.}
	\label{fig:setupCu}
\end{figure}

\begin{figure}
	\includegraphics[width=\columnwidth]{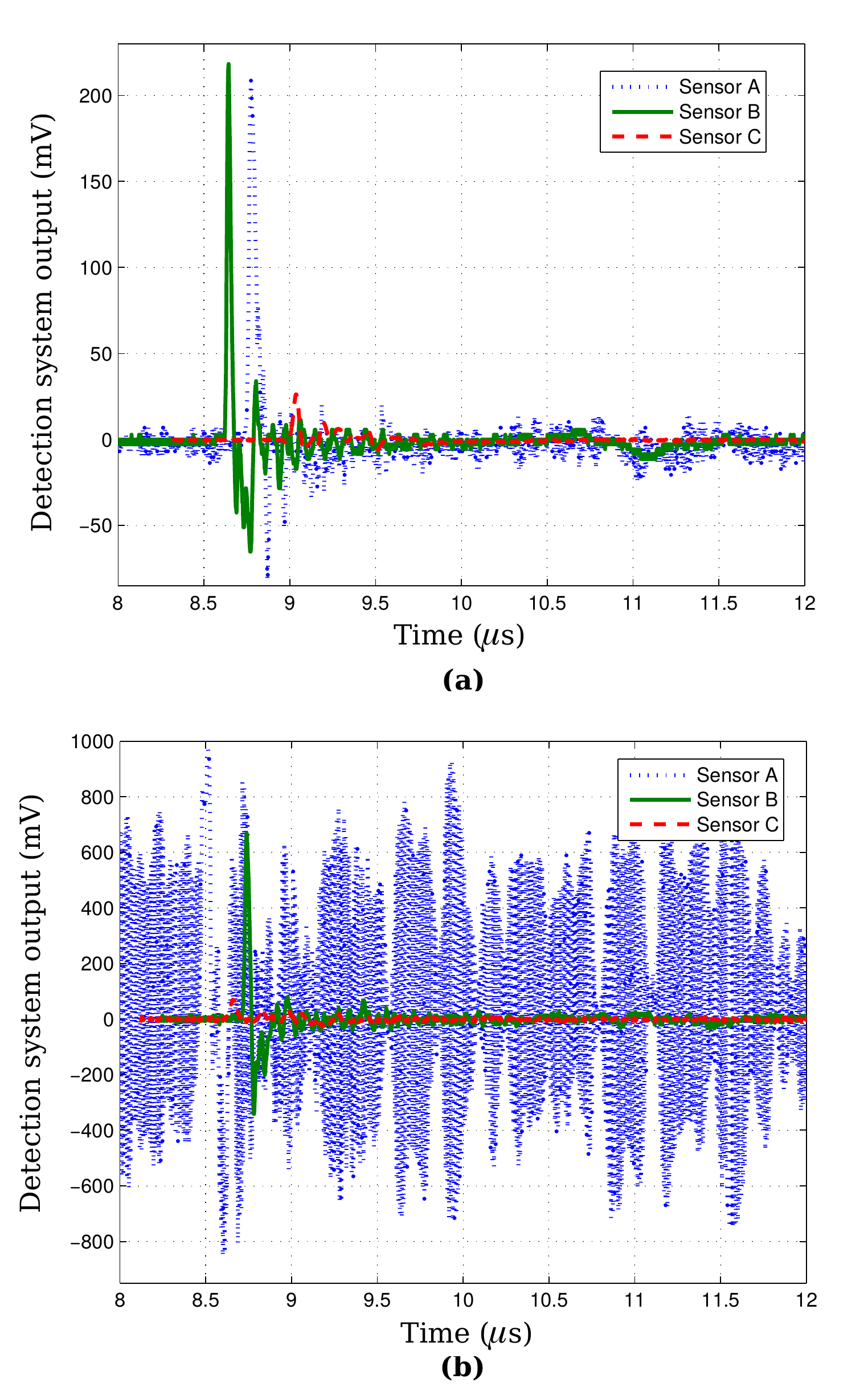}
	\caption{Detection system response for a pressure pulse (10 ns) with a peak value of 485 Pa. (a) Sensor + TIA1. (b) Sensor + TIA2}
	\label{fig:STIA1y2}
\end{figure}

The acoustic characterization of the detection system was performed following the method described in ref. \cite{gonzalez2019}. Fig.~\ref{fig:setupCu} presents the experimental setup. The piezoelectric sensor and a metallic target, used as the acoustic source, were immersed in a cylindrical vessel (115 mm radius x 150 mm height) filled with deionized water. A frequency-doubled Nd:YAG laser (Continuum Minilite I, 532 nm, 5 ns, 10 Hz) and a converging lens irradiated a 100 $\mu$m diameter copper wire. The focused laser beam had a spot diameter close to that of the wire, thus providing a roughly spherical irradiated volume. The illuminated wire generates sub-microsecond quasi-unipolar pressure \cite{gonzalez2019}. Two XYZ translation stages adjusted the position of the acoustic source and the piezoelectric sensor, respectively. The laser Q-switch pulse triggered a Tektronix TDS 2024 oscilloscope (2 GS/s, 200 MHz), that digitized the output of the transimpedance amplifier. A pyroelectric detector (Coherent J-10MB-LE) measured the laser pulse energy. 

Fig.~\ref{fig:STIA1y2} shows the response of the detection system with TIA1 and TIA2 for a pressure pulse with a peak value of 485 Pa and a duration of 10 ns. In Fig. ~\ref{fig:STIA1y2} (a), it can be seen that sensors A and B have a similar sensitivity whereas sensor C has a much lower value. This is due to the silver paint layer deposited on the detection active zone that causes a mismatch of acoustic impedance. Assuming plane waves, the reflection coefficient between water ($c$=1500 m/s, $\delta$= 1 g/cm$^3$) and silver paint ($c$=2550 m/s, $\delta$= 6.5 g/cm$^3$) is greater than 0.8. 

The plots in Fig.~\ref{fig:STIA1y2} (b) show that the detection systems with TIA2 have a higher sensitivity but the combination sensor A + TIA2 has a very poor signal to noise ratio (SNR $<$ 0 dB). The SNR was calculated as the ratio of instantaneous signal power to noise variance \cite{telenkov2013}.  

\subsection{Noise equivalent pressure}

\begin{figure}
	\includegraphics[width=\columnwidth]{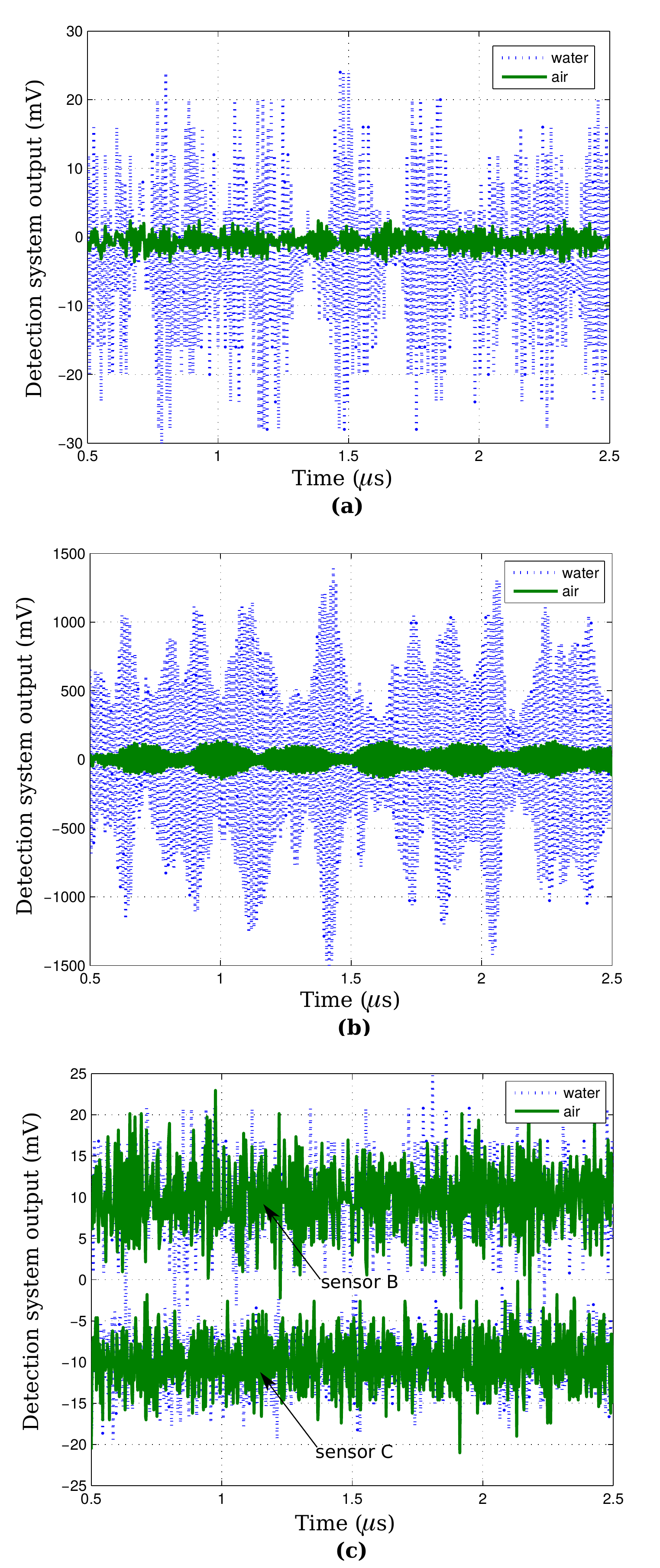}
	\caption{Recorded noise signal for two different acoustic coupling media: air (green solid line) and water (blue dashed line). (a) Sensor A + TIA1. (b) Sensor A + TIA2. (c) Sensor B (or C) + TIA2}
	\label{fig:NAA}
\end{figure}

The noise equivalent pressure NEP (noise floor) of the detector is an important figure of merit that can be expressed as a spectral density in units of Pa/$\sqrt{\textnormal {Hz}}$ \cite{winkler2013}. In OAT, a low-noise, large bandwidth detection system allows obtaining high resolution images (small size objects). In order to study the main noise source of the implemented detectors, we repeated the same measurements presented above but using air as an acoustic coupling medium. This way, the captured noise is mostly electrical. Fig.~\ref{fig:NAA} shows the signals using deionized water (dashed blue line) and air (solid green line) in a time interval before the acoustic pulse reaches the detector. Curiously, the noise with the detector immersed in water has always larger amplitude than in air. These results suggest the electrical properties of water play an important role. In the case of sensor A, the difference of noise amplitude between air and water is higher than the obtained values with sensors B and C. Moreover, in detector C, the difference is almost negligible. The behavior of sensors B and C with TIA1 is similar to that shown in Fig. \ref{fig:NAA} (c).

\begin{table*}
	\centering
	\begin{tabular}[t]{@{}lllllll}
		\hline
		&  \multicolumn{2}{c}{Sensor A} & \multicolumn{2}{c}{Sensor B} & \multicolumn{2}{c}{Sensor C} \\
		\hline
		& TIA1  & TIA2  & TIA1 & TIA2 & TIA1 & TIA2 \\
		\hline
		Sensitivity ($\mu$V/Pa) & 430  & 1690 & 450 & 1630 & 54 & 145 \\
		
		Bandwidth (MHz) & 20  & 63 & 19 & 62 & 20 & 60 \\
		
		NEP (mPa/$\sqrt{\textnormal {Hz}}$) in water & 5.4  & 33 & 0.4 & 3.0 & 1.5 & 2.8 \\
		
		NEP (mPa/$\sqrt{\textnormal {Hz}}$) in air & 0.8  & 3.8 & 0.3 & 0.2 & 1.3 & 2.5 \\
		\hline
	\end{tabular}
	\caption{Main results of the sensors characterization. NEP: Noise Equivalent Pressure single shot value.}
	\label{t:resul}
\end{table*}

Table \ref{t:resul} presents a summary of the main results obtained in this section. The detection system sensitivities were calculated as the peak voltage of the OA signal divided by the peak value of the detected pressure pulse generated by the copper wire. The NEP was determined as the ratio of noise variance to the sensitivity and the square root of the system bandwidth. The detection system integrated by sensor B with TIA2 has high sensitivity, large bandwidth and low noise equivalent pressure (NEP), in comparison with the other detection systems studied in this work. Moreover, the best NEP value (detector B + TIA2) is similar or better than other integrating piezoelectric transducers reported in the literature \cite{nuster2009,winkler2013,paltauf2017}.  

Finally, so as to verify the repeatability of the results, we implemented three more sensors with the characteristics of type B and we obtained similar results ($\pm$ 10 $\%$).

\subsection{Dielectric resonator antenna effect}

\begin{figure}
	\includegraphics[width=\columnwidth]{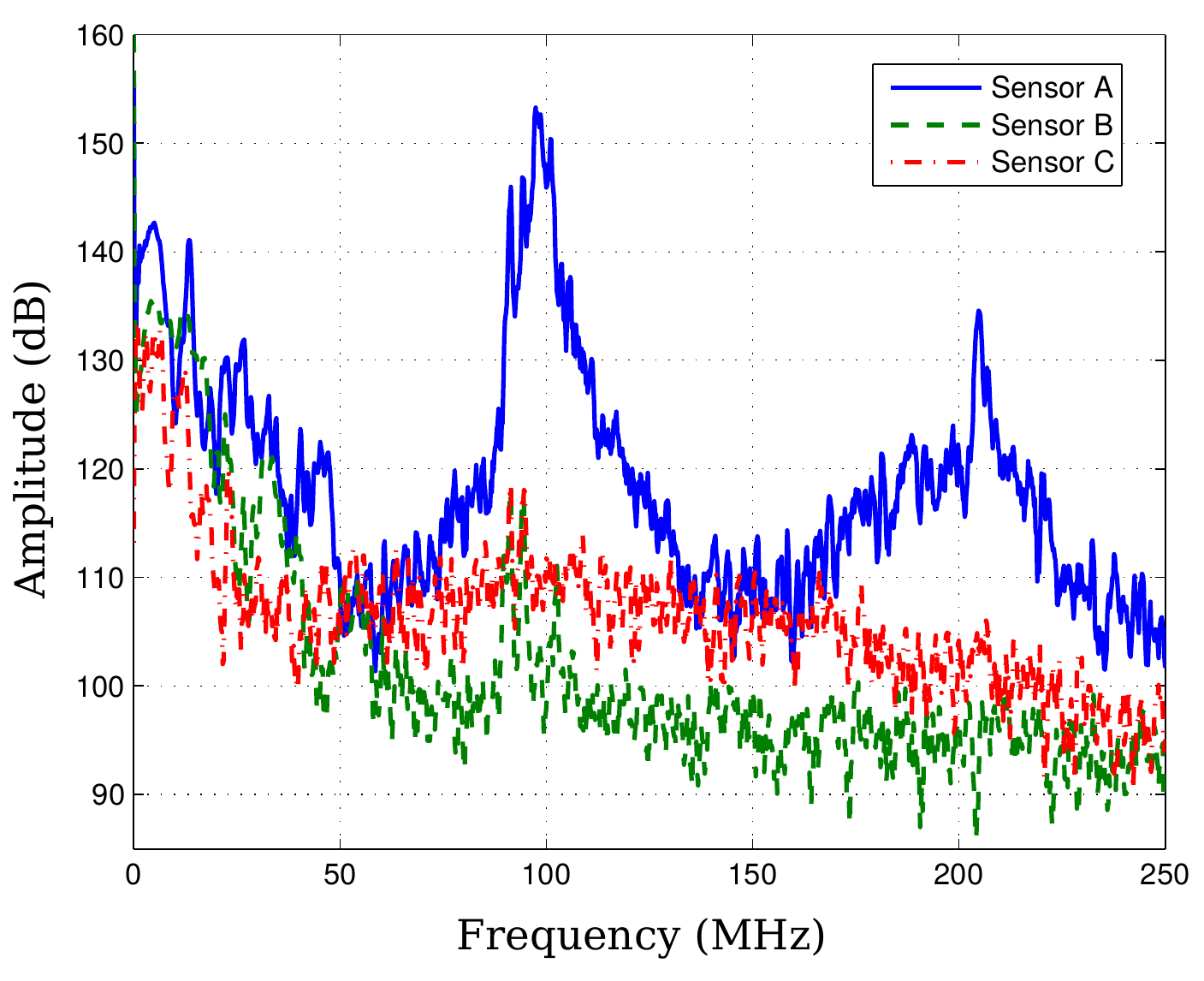}
	\caption{Fourier spectrum of signals presented in Fig. \ref{fig:STIA1y2} (b). The sound pressure reference level is 1 $\mu$Pa.}
	\label{fig:spectrum}
\end{figure}

Fig. \ref{fig:spectrum} depicts the Fourier spectrum of the signals presented in Fig. \ref{fig:STIA1y2} (b). It can be seen that the spectrum obtained with sensor A + TIA2 in water shows a strong peak around 100 MHz. This phenomenon occurs since our detector and the vessel filled with water resemble a dielectric resonator antenna (DRA). The basic principle of operation of dielectric resonators is similar to that of the cavity resonators \cite{pozar2012} and is thoroughly discussed in the literature \cite{keyrouz2016}. The typical antenna consists of a cylindrical dielectric resonator (DR) with height $h$, radius $a$, and dielectric constant $\epsilon_{r}$. The DR is placed on top of a ground plane and fed by a coaxial cable. In our case, the DR is the vessel filled with water, and the coaxial cable is the integrating line sensor. The resonant frequencies of the modes supported by a cylindrical DRA can be computed as \cite{Luk2003}:

\begin{equation}
f_{TEnpm} = \frac{c}{2\pi a \sqrt{\epsilon_r \mu_r}}\sqrt{X_{np}^2+\left(\frac{\pi a}{2h}(2m+1)\right)^2}
\label{fTE}
\end{equation} 

\begin{equation}
f_{TMnpm} = \frac{c}{2\pi a \sqrt{\epsilon_r \mu_r}}\sqrt{X_{np}^{\prime 2}+\left(\frac{\pi a}{2h}(2m+1)\right)^2}
\label{fTM}
\end{equation}

\noindent
where $X_{np}$ and $X'_{np}$ are the roots of the Bessel functions and first-order derivative Bessel functions of the first kind, respectively. The resonant frequency decreases if the radius, the height or the dielectric constant of the DRA increase. This behavior is the most important characteristic of the DRA since it allows reducing the size of the DRA by employing a higher dielectric constant. It is important to note that, when the probe length ($L$) (sensor length) is comparable with the dimensions of the DR ($h$ or $a$), the input impedance of the DRA can be tuned and, consequently, the resonance frequency can be controlled \cite{leung1993}. The longer the probe length, the lower the resonant frequency. However, when $L << h,a$ (our case), small changes in $L$ do not introduce a sizable change on the value of the resonant frequency \cite{junker1994}. With the dimensions of our vessel and considering water at 20$^o$C ($\epsilon_r = 80$ and $\mu_r = 1$ \cite{malmberg1956}), we calculated a resonance frequency of the fundamental mode TM$_{110}$ of $\sim$100 MHz; a result very close to the measured noise signal (see Fig. \ref{fig:spectrum}). This value is quite inconvenient because many FM stations populate the electromagnetic spectrum in the 88-108 MHz range. Therefore, good electrical shielding is mandatory if low NEP is sought. It can be seen, in Figs. \ref{fig:NAA} (c) and \ref{fig:spectrum}, how the effects of this phenomenon can be significantly reduced ($\sim$ $-$50 dB at 100 MHz) by covering the detector with a thin layer of silver paint. In our case painting is simpler than bending and fitting a metal sheet around the detector and silver has the highest electrical conductivity and thus, the smallest skin depth at radio frequencies. 

It is important to note that it is possible to shift the resonance frequencies to higher values by reducing the radius or height of the vessel. In fact, we tested another vessel ($a=$ 50 mm, $h=$ 100 mm) obtaining a resonance frequency of $\sim$400 MHz and a noticeable decrease in noise. However, the object to be imaged must be smaller, limiting the applications of the OAT system in biological imaging. A solution to the DRA problem might be doing away with water, but this is not easy because it implies finding a low $\epsilon_r$ liquid whose acoustic impedance is close to that of water (best choice for biological samples). In addition, it should also be chemically inert. Finally, filtering the recorded signal is not a solution because, if the noise level is too high, the acquisition system might not have enough bits to accurately register the OA signal. 

\section{Application to OAT}
\label{s:toa}

\begin{figure}
	\includegraphics[width=\columnwidth]{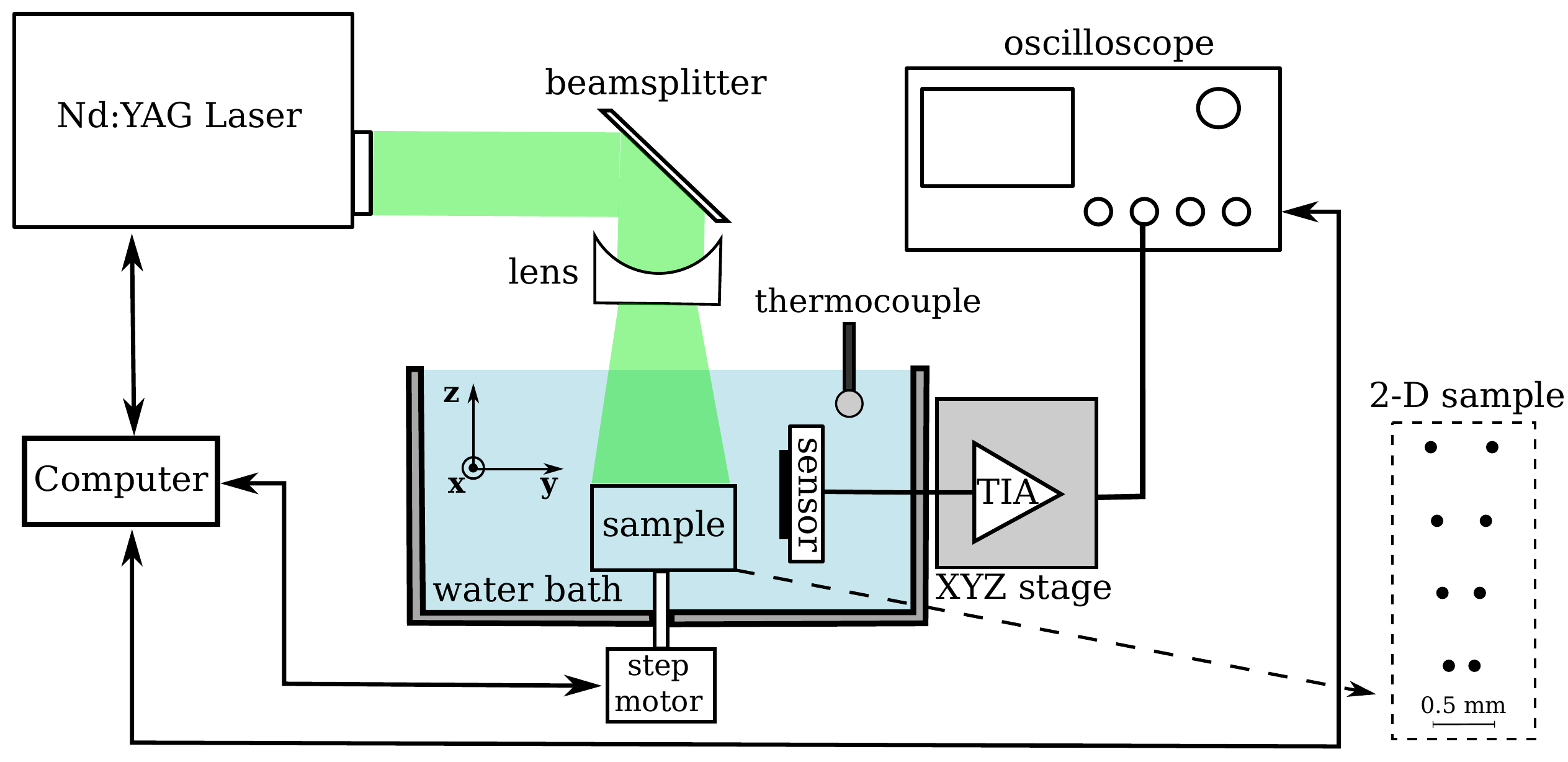}
	\caption{OAT system scheme used to assess the detection system performance.}
	\label{fig:setupTOA}
\end{figure}

In order to compare the performance of the detection systems I (sensor A + TIA1) and II (sensor B + TIA2), we used them in the 2-D OAT described in ref \cite{gonzalez2018}. Fig.~\ref{fig:setupTOA} shows the experimental setup scheme. The sensor and the sample were immersed in a vessel filled with deionized water (same size of the previous setup). The water temperature was measured with a calibrated thermocouple to determine the speed of sound. A Nd:YAG laser with second harmonic generation (Continuum Minilite I, 532 nm), 5 ns pulse duration, 10 Hz repetition rate and pulse energy less than 1 mJ, was the light source. A diverging lens adapted the diameter of the laser beam to a size larger than the sample, trying to achieve homogeneous illumination. The ultrasonic detector was fixed and pointed to the center of the rotating sample stage using an XYZ translation stage. Phantoms were fixed to a rotatory stage (Newport PR50CC) and rotated 360$^{\circ}$ in 1$^\circ$ steps since full view data (i.e. 360$^\circ$) minimizes the effect of a limited view detection \cite{xu2005}. The OA signals were recorded and averaged 64 times. 

\begin{figure}
	\includegraphics[width=\columnwidth]{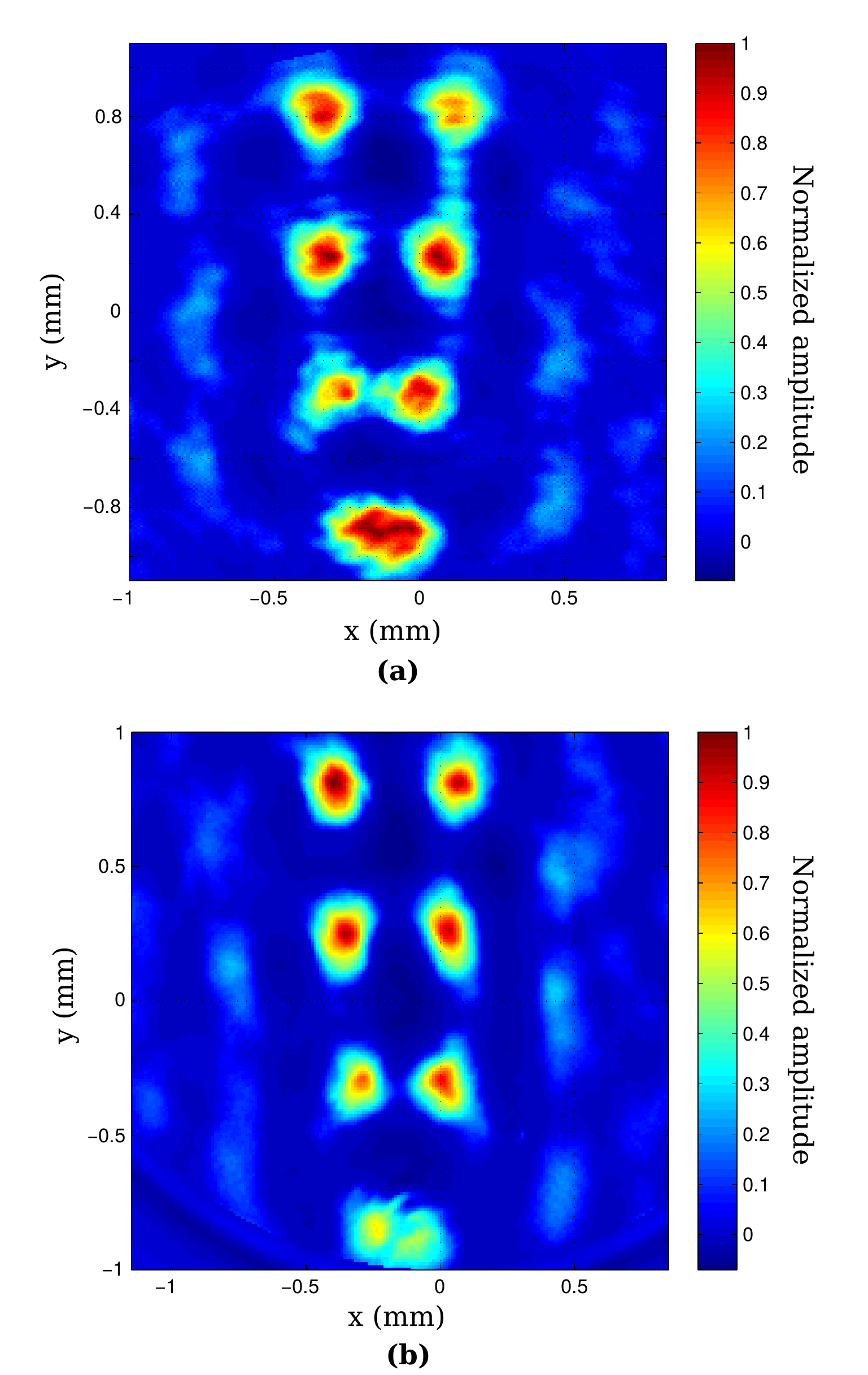}
	\caption{Reconstructed image for the detection system integrated by (a) sensor A and TIA1 and (b) sensor B and TIA2.}
	\label{fig:OATVyN}
\end{figure}

The sample consists of a transparent film embedded in agarose gel. A laser printer drew on the film a pattern of black disks (100 $\mu$m of diameter) at different distances (see Fig. \ref{fig:setupTOA}). The agarose gel was prepared with 2.5$\%$ (w/v) agarose in distilled water. First, a cylindrical base of the agarose gel with a diameter of 15 mm and a height of approximately 25 mm was prepared. Then, the object (transparent film) was placed in the middle of the cylinder and fixed with a few drops of gel. Finally, another layer of gel ( 1 mm thickness) was deposited on top of the sample object. 

The projections of the initial pressure distributions into the $xy$ plane were obtained using the backprojection algorithm described in ref. \cite{burgholzer2007}.

Fig. \ref{fig:OATVyN} presents the reconstructed images for the detection systems I and II. As expected, the detection system with lower noise and larger bandwidth distinguishes better between two very close black disks and has a higher level of detail at the borders of the disks of the image (higher acuteness). 

To assess the system's resolution, we compared the full width at half maximum (FWHM) of the intensity profiles between the theoretically printed ink pattern and the reconstructed image along the $x$ and $y$ axis. We repeated this procedure on 10 identical phantoms obtaining 40 values of FWHM for each axis. The resolutions of the systems I and II are 195 $\pm$ 10 $\mu$m and 100 $\pm$ 10 $\mu$m, respectively. Considering the finite size of the printed patterns, this can be regarded as an upper limit of the resolution \cite{paltauf2017}. 

\section{Conclusions}
\label{s:conclu}

In this work we studied the origin of a high-noise floor in a water-immersed, integrating-line piezoelectric sensor, and a way of reducing it. We found that the detector, in conjunction with the acoustic coupling medium (water), behaves as a resonant dielectric antenna with resonance frequencies close to those of interest. This phenomenon limits the performance of the system because it helps capturing unwanted signals, whose frequency is in the range of the desired ones. We achieved good shielding, and lower noise equivalent pressure, by using silver-based paint that fits snugly around the sensor and does not degrade the acoustic sensitivity. Alternative solutions, such as reducing the size of the vessel, replacing water with another liquid or filtering the recorded signal were discarded because of the increased complexity. The new sensor has improved noticeably. This allowed the efficient use of a transimpedance amplifier with high gain and large bandwidth. The result was a broadband detection system with high sensitivity and low equivalent noise pressure. The improved system offers an easy-to-implement, highly repeatable and low-cost option that is suitable for home-made tomographic devices. 

\section*{Acknowledgment}

This work was supported by the Agencia Nacional de Promoci\'on Cient\'ifica y Tecnol\'ogica (PICT Grant No. 2016-2204).

\section*{References}

\bibliographystyle{iopart-num.bst}
\bibliography{Insa_references.bib}

\end{document}